\RequirePackage{fix-cm}
\documentclass[smallextended]{svjour3}
\smartqed  
\usepackage[T1]{fontenc}
\usepackage{amsmath}
\usepackage{mathptmx}
\usepackage{graphicx}
\usepackage{url}
\usepackage{natbib} 
\usepackage{subfig}
\usepackage{parskip}

\journalname{Experimental Astronomy}
\begin{document}
\title{The SFXC software correlator for Very Long Baseline Interferometry: Algorithms and Implementation}
\titlerunning{The SFXC software correlator for VLBI: Algorithms and Implementation}
\author{A. Keimpema \and M.M. Kettenis \and S.V. Pogrebenko \and 
R.M. Campbell \and
G. Cim\'o \and
D.A. Duev \and
B. Eldering \and
N. Kruithof \and
H.J. van Langevelde \and
D. Marchal \and
G. Molera Calv\'es \and
H. Ozdemir \and 
Z. Paragi \and
Y. Pidopryhora \and
A. Szomoru \and
J. Yang
}
\institute{A. Keimpema, M.M. Kettenis, S.V. Pogrebenko, R.M. Campbell, G. Cim\'o,  
D.A. Duev, B. Eldering, H.J. van Langevelde, G. Molera Calv\'es, Z. Paragi, A. Szomoru
\at
           Joint Institute for VLBI in Europe \\
           Oude Hoogeveensedijk 4 \\
           7991 PD Dwingeloo \\
           Tel.: +31 521 596 524
           \and
           H.J. van Langevelde \at
           Also affiliated with\\
           Sterrewacht Leiden, Leiden University\\
           P.O. Box 9513, 2300 RA, Leiden
           \and
           N. Kruithof \at
           Cruden B.V.\\
		   Pedro de Medinalaan 25, 1086 XP Amsterdam
		   \and
		   D. Marchal\at
		   Laboratoire d'Informatique Fondamentale de Lille (LIFL), University of Lille\\
		   Cit\'e scientifique - B\^atiment M3\\
		   9655 Villeneuve d'Ascq C\'edex
		   \and
		   H. Ozdemir\at
		   Energieonderzoek Centrum Nederland (ECN)\\
		   P.O.Box 1, 1755 ZG Petten
		   \and
		   Y. Pidopryhora\at
		   School of Mathematics and Physics, University of Tasmania,\\
		   Private Bag 37, Hobart, Tasmania 7001, Australia
		   \and
		   J. Yang\at
		   Department of Earth and Space Sciences, Chalmers University of Technology\\
		   Onsala Space Observatory, SE-43992 Onsala, Sweden
}
\date{Received: date / Accepted: date}
\maketitle

\begin{abstract} 
In this paper a description is given of the SFXC software
correlator, developed and maintained at the Joint Institute for VLBI
in Europe (JIVE). The software is designed to run on generic
Linux-based computing clusters. The correlation algorithm is explained
in detail, as are some of the novel modes that software correlation
has enabled, such as wide-field VLBI imaging through the use of
multiple phase centres and pulsar gating and binning. This is followed
by an overview of the software architecture. Finally, the performance
of the correlator as a function of number of CPU cores, telescopes and
spectral channels is shown. 
  \keywords{VLBI \and radio astronomy \and software correlation} 
\end{abstract}

%
\section{Introduction}
%
Very Long Baseline Interferometry (VLBI) is a technique in which the signals from a
network of radio telescopes, spread around the world, are combined to create one single
telescope with a resolution far surpassing that of the individual telescopes. This is
accomplished by computing the correlation functions on every baseline, formed by all
possible pairs of telescopes, on a central data processor called the correlator. The
digitised signals are recorded on magnetic media which are shipped to the correlator, or,
with the advent of real-time electronic VLBI~(e-VLBI)~\citep{Szomoru06}, streamed directly over the Internet.
From its inception in 1980~\citep{Porcas10}, the European VLBI Network (EVN) has relied on
correlators built from dedicated ASIC-based hardware, because the processing of VLBI
observations far exceeded the capacity of standard computing components (see e.g.
\citet{Schilizzi01}). However, the continuously increasing computing power of modern
CPUs has made it possible to use commercial-of-the-shelf (COTS) components for
correlation. Together with the increase in storage capacity this has changed the way VLBI
is done.

Development of the Super FX Correlator (SFXC) at JIVE started with the purpose to detect and track a
spacecraft in the outer solar system, namely ESA's Huygens Probe as it descended to the
surface of Saturn's moon Titan~\citep{Pogrebenko04}. Such an application of VLBI calls
for extremely high spectral resolution and a detailed control of the delay model, both of
which were in practice not available in existing hardware correlators.
From then on, the SFXC software correlator has been further developed into a very
efficient and flexible system, providing a number of completely new observing modes for the EVN. 

The introduction of software correlators has brought about several
scientific improvements. Although 2-bit representation of the data is still used for
economy of storage and transport, the algorithm used for correlation is no longer limited
to few-bit precision. This means that the data can be processed with much higher accuracy.
In contrast to hardware correlators, software correlators can operate asynchronously,
certainly when the data are read from disk systems. The consequence of the above changes
is that correlation functions can be obtained with almost arbitrary windowing functions
and accumulation times. This results in a much extended coverage of parameter space,
providing for example arbitrarily fine spectral resolution and very flexible pulsar gating
and binning.

In addition, the software correlator has far more flexible interfaces to data sources,
geometric model components and data output, as these are now entirely implemented in
software. The software correlator is therefore a natural platform to process
real-time e-VLBI data. The flexible interfaces also make it
possible to cross-correlate different data formats, compensate for problems with recording times
and implement multiple output streams for more than one target position within
the field of view of the observations \citep[e.g.][]{Cao14}.

The software correlator runs on a standard Linux cluster and has become the operational
correlator for all EVN operations at JIVE. SFXC is available under version 2 of the 
GNU Public License~(GPL)\footnote{\url{http://www.gnu.org/licenses/gpl-2.0.html}}.
It has proven to be a very stable platform, routinely producing high-
quality scientific results for the EVN community. In almost all aspects 
it compares favourably to the old MkIV EVN hardware correlator~\citep{Schilizzi01}.
It should finally be noted that power consumption is higher for software correlators, this
situation will improve with time as the drive towards green computing produces ever more
efficient processors.

A number of other VLBI software correlators have been developed over
the past decade. Some examples are the K5 software correlator, of the
Japanese National Institute of Information and Communications
Technology (NICT)~\citep{Kondo04}, used mainly for geodetic
applications, the SOFTC correlator~\citep{Lowe04}, which is the
operational VLBI correlator of the Jet Propulsion Laboratory (JPL),
and the DiFX correlator~\citep{Deller07,Deller11}, maintained by an
international consortium of developers and used at a number of sites
such as the Very Long Baseline Array (VLBA), Australian Long Baseline Array (LBA), and the
Max Planck Institute for Radio Astronomy (MPIfR).

\section{Correlation algorithm}
\label{sec:coralg}
%
\begin{figure}
  \centering
  \includegraphics[width=0.8\linewidth]{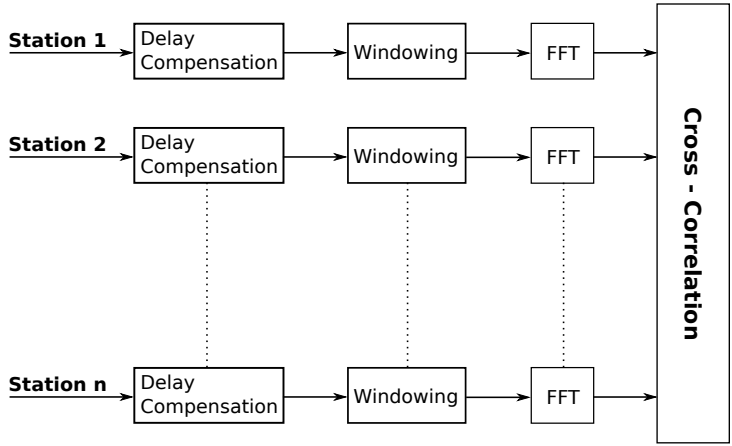}
  \caption{The data flow inside a correlator node.}
  \label{Fig:coralg}
\end{figure}
In this section we present the correlation algorithm implemented in SFXC.
We will limit the discussion to our design considerations. A more in-depth review 
of radio interferometry and VLBI can be found in standard references like e.g.
~\citet{Thompson01} and~\citet{Taylor99}. In Fig.~\ref{Fig:coralg} the
data flow inside the correlator core is shown and in the following subsections we shall
elaborate on each step in this flow. 
%
\subsection{Delay compensation}
\label{sec:delaycomp}
%
Prior to cross correlation the signals from each antenna must be brought to a common
frame of reference. For most applications this is the geocentric frame. This is
accomplished by time-shifting the signals from each antenna such that at all signals 
track the same wave front. The default delay model used in SFXC is the
CALC 10 software\footnote{\url{http://gemini.gsfc.nasa.gov/solve}} developed at the Goddard 
Space Flight Center. This is the same model that was
used for the EVN MkIV correlator at JIVE~\citep{Schilizzi01}.
The default delay model can overridden by a user supplied model. For instance, 
\citet{Duev12, Duev14} developed a near-field delay model for spacecraft tracking
applications in combination with SFXC and a delay model that has been successfully used to
correlate data observed with the RadioAstron orbital telescope.
Before correlation the delay model is evaluated at one second intervals,
which allows the model to be interpolated close to the machine precision~\citep{Ozdemir}.
During correlation these values are interpolated using an Akima spline~\citep{Akima70}. 
The Akima spline was chosen because it yields accurate results also at the edges of the interpolation interval.

Compensation for the geometric delay $\tau$ is performed in two steps. In the first step
the input stream is shifted by an integer number of samples. For a Nyquist sampled
signal the corresponding delay $\tau_i$, called the integer delay, rounding to the nearest
integer,
\begin{equation}
\tau_i = \mathtt{round}(2B\tau)/2B,
\end{equation}
with $B$ the sampled bandwidth.
The residual fractional delay, given by 
\begin{equation}
\tau_f = \tau - \tau_i,
\end{equation}
is at most half a sample period. The fractional delay leads to a phase error across
the band 
\begin{equation}
\phi_{\mathrm{frac}}(\nu) = 2\pi\tau_f\nu,
\label{eq:frac_delay}
\end{equation}
where $\nu$ is the baseband frequency. In order to remove the fractional delay error, the data is
divided into segments of $N$ samples, each segment is Fourier transformed, and the resulting
spectra are multiplied by $e^{-i\phi_\mathrm{frac}}$. We will discuss the choice of the segment
length $N$ later in this section.

The astronomical signal as it is received by a telescope is down converted from a frequency range 
$\nu_{sky}\leq \nu <\nu_{sky}+B$ to a range $0\leq\nu<B$, the so-called baseband.
Since the fractional delay compensation is performed on the base band signal a phase error
$\phi_\mathrm{frot}$ will remain:
\begin{equation}
\phi_\mathrm{frot}=2\pi\nu_\mathrm{sky}\tau.
\label{eq:coralg-gamma}
\end{equation}
The effect of $\phi_\mathrm{frot}$ is most easily shown by looking at a single baseline.
Let $\tau_b$ be the delay between the two telescopes on that baseline. 
If $\phi_\mathrm{frot}$ is not corrected, the fringe amplitudes on this baseline will
oscillate with a frequency $\nu_{\mathrm{sky}}\textrm{d}\tau_b/\textrm{d}t$. This effect
is removed by multiplying the time domain baseband voltage data for each antenna with 
$e^{-i\phi_\mathrm{frot}}$, a process which is called fringe rotation or fringe stopping.

If the delay $\tau$ were constant then the combination of delay compensation and
fringe rotation would exactly compensate for the difference in arrival time $\tau$. In the 
general time dependent case this correction is only exact for a single frequency
$\nu_\mathrm{frot}$, which we call the fringe rotation frequency. In Eq.~\eqref{eq:frac_delay} and Eq.~\eqref{eq:coralg-gamma} 
the fringe rotation frequency is at the band edge, $\nu_\mathrm{frot} = \nu_\mathrm{sky}$. The 
fringe rotation frequency can be shifted by an arbitrary amount $\nu_\delta$ by applying a phase shift
\begin{equation}
\psi(\nu_\delta) = e^{-i2\pi\nu_\delta\tau}.
\end{equation}
which will shift the fringe rotation frequency to
$\nu_\mathrm{frot}=\nu_\mathrm{sky}+\nu_\delta$. The phase shift that should be applied in the
fringe rotator then becomes
\begin{equation}
\phi_\mathrm{frot}(\nu_\delta)=2\pi(\nu_\mathrm{sky}+\nu_\delta)\tau=2\pi\nu_\mathrm{frot}\tau.
\label{eq:coralg-gamma2}
\end{equation}
In principle $\nu_\delta$ could be chosen to align with a spectral line to maximize the signal to
noise at this spectral point. In SFXC we set $\nu_\delta$ equal to $B/2$ which moves
$\nu_\mathrm{frot}$ to the centre of the band, minimizing the average error.

The act of delay compensation and fringe rotation is equivalent to applying a Lorentz
transformation, transforming the signal from the reference frame of each telescope 
to that of a fictional observer at the geocentre. At this point the data can be 
re-analysed at an arbitrary spectral resolution, unrelated to that used in the delay
compensation. This decoupling of spectral resolution in the delay compensation and
cross-correlations is essential to achieve true arbitrary spectral resolution.

In the fractional delay correction a single delay value is used for an entire
FFT segment, which constrains the FFT period to some fraction of the time between two
integer delay changes. On short time scales the delay $\tau(t)$ is approximately linear
\begin{equation}
\tau(t) \approx \tau_0 + (t-t_0)\dot{\tau},
\end{equation}
where $\tau_0$ is the delay at the time $t_0$, and $\dot\tau$ is the delay rate. The number of
samples $N$ between two integer delay changes is independent of bandwidth and is given by
\begin{equation}
N = \dot\tau^{-1}.
\end{equation}
Following the discussing in Sec.~9.7 of~\citet{Thompson01}, the drift in delay during the FFT 
period will cause a loss in amplitude of,
\begin{equation}
  L(\nu) = \frac{\sin({\pi(\nu-B/2)\dot{\tau}T})}{\pi(\nu-B/2)\dot{\tau}T},
\end{equation}
where $\nu$ is the baseband frequency, $B$ the bandwidth, and $T$ the duration of the FFT.
Additionally, there will be a negligible phase error proportional to $\ddot\tau T^2$.
On terrestrial baselines the maximum geometric delay rate that can be encountered is 
$\dot\tau_t = 1.5\mu s/s$. 
In order to keep the drift in delay over one FFT segment
within 0.1 samples, the maximum FFT size, rounding down to a power of two,  is \
$N_{max} = 65536$ samples. For near-field VLBI the delay rates can be considerably higher.
For example satellites in a low earth orbit move at velocities of up to 8 km/s, which translates 
to a delay rate of $\dot\tau=27\mu s/s$. For such satellites the maximum FFT window size
becomes $N_{max}=2048$.
%
\subsection{Windowing} 
\label{sec:windowing}
%
An FX correlator, such as SFXC, computes a cyclical correlation function rather than a linear
correlation. This is a direct consequence of the discrete correlation theorem 
\begin{equation}
\Gamma[m] =
\sum_{n=0}^{N-1}f[(n+m)\textrm{mod}N]g^*[n]= 
\sum_{k=0}^{N-1}F[k]G^*[k]e^{i2\pi km/N},
\end{equation}
where $N$ is the FFT size and $F[k](G[k])$ is the Fourier pair of $f[n](g[n])$. 
By applying an appropriate windowing function this cyclicity can be reduced or even removed
completely.

At the windowing stage, the data stream of each telescope is divided into segments of $N$ samples
which are then multiplied by a windowing function. These windowed segments are then Fourier
transformed and the cross-correlations product are computed. Consecutive segments are 
half-overlapped. 

The cross-correlation function after applying a windowing function $w[n]$ becomes
\begin{equation}
\Gamma[m] = (w[n]\cdot f[n])\ast (w[n]\cdot g^*[n])=
            \mathcal{F}^{-1}\left(W^2[k]\ast (F[k]\cdot G^*[k])\right),
\end{equation}
where $W[k]$ is the Fourier pair of $w[n]$. This means that the spectrum is convolved with 
the square of the Fourier transform of the windowing function.

The choice of window function is a compromise between spectral leakage effects and
spectral resolution. E.g. the square window defined by
\begin{equation}
w[n] = \left\{\begin{array}{l l l} 1 & & n<=N/2\\ 0 & & n > N/2 \end{array}\right.
\end{equation}
convolves the spectrum of the cross-correlation function with a $\textrm{sinc}^2$ function. 
The sinc function has a narrow main lobe and therefore excellent resolution, but large slowly
decaying side lobes which can potentially obscure spectral line features. This ringing effect is 
due to the hard edge at $n = N/2$. Smoother windows, such as the Hann window, will reduce leakage
effects but at the cost of resolution. The loss in resolution can be compensated by increasing the
number of spectral points, but this in turn leads to larger output data volumes. The windowing 
functions that are currently available in SFXC and their most important characteristics are listed
in Table~\ref{tab:coralg-windows}. 

\begin{table}
\begin{tabular}{|l|l|c|c|}
\hline
Window name & Definition & Main lobe FWHM & First side lobe\\
\hline
Square & $w[n] = 1$ if $n<=N/2$ else $0$ & 1.21 & -13 dB\\
Cosine & $w[n] = \cos(\pi n/(N-1))$ & 1.64 & -23 dB\\
Hann & $w[n] = 0.5 (1 - \cos(2\pi n/(N-1))$ & 2.00 & -31 dB\\
Hamming & $w[n] = 0.54 - 0.46\, \cos(2\pi n/(N-1))$ & 1.82 & -43 dB\\
\hline
\end{tabular}
\caption{Available window functions and their characteristics}
\label{tab:coralg-windows}

\end{table}
%
\subsection{Cross-correlation and normalization} 
%
In the final step auto- and cross-correlations are computed for all
baselines. Whether to compute cross-hand polarisations or only
parallel polarisations is a user configurable option. The requested correlation products are
accumulated in a buffer until the end of the integration period after which they are written 
to disk.
The correlator integration time is defined according to the scientific goals of the
observations. For regular VLBI projects it is typically 1-4s, but can be much shorter
e.g. for wide field of view applications.

At the end of the integration period, the visibilities are normalized. 

According to the Wiener-Khinchin theorem the total power contained in a signal 
can be obtained from the auto-correlations.
If $P_i$ is the total power of telescope $i$, the normalisation factor for 
baseline pair $(i,j)$ will be
\begin{equation}
A_{ij} = \sqrt{P_iP_j}.
\label{eq:norm}
\end{equation}
In practice the computation of the normalisation factor is complicated by the fact that
a certain fraction of the input samples may be absent. This for example can be caused by
a faulty hard drive or missing network packets. Furthermore, some data formats use data 
replacing headers and therefore will have missing samples even in the absence of any disk 
or network failures. Missing samples are flagged and replaced with zeros 
and will therefore not contribute to the cross-correlations. If during some period of time
one telescope has invalid data but the other does not then Eq.~\eqref{eq:norm} will 
overestimate the cross-spectrum power.
The normalization factor should be scaled by an additional factor 
\begin{equation}
C_{ij} = N^{(ij)}_{\textrm{bl}} / N^{(i)}_{\textrm{valid}},
\end{equation}
where $N^{(ij)}_{\textrm{bl}}$ is the number of samples for which both telescopes on the
baseline simultaneously have valid data, and $N^{(i)}_{\textrm{valid}}$ is the total number of
valid  samples for telescope $i$. The normalisation factor then becomes
\begin{equation}
A_{ij} = \sqrt{C_{ij}C_{ji}P_iP_j}.
\label{eq:norm2}
\end{equation}
%
\section{Pulsar binning} 

In this section we will discuss the various aspects relating to the correlation of 
pulsar data. A fundamental property of pulsars is that although their individual pulses
can have widely varying pulse shapes and arrival times, the long term average pulse
profile is highly stable~\citep{Lorimer05}.
Due to this property high precision models of the pulsar rotational phase can be created.
For many pulsars the duty cycle, $\tau_d$, defined as the ratio between pulse width and pulse
period, is relatively small. A significant improvement in signal to noise for VLBI observations
can be achieved by accumulating the correlation function only during pulse reception. This
is commonly referred to as pulsar gating. Using this approach the signal to noise can be
increased by approximately $\sqrt{1/\tau_d}$, which typically is a factor of 3-5. 

\begin{figure}
 \centering
 \subfloat[(a)][]{ \includegraphics[width=0.45\linewidth]{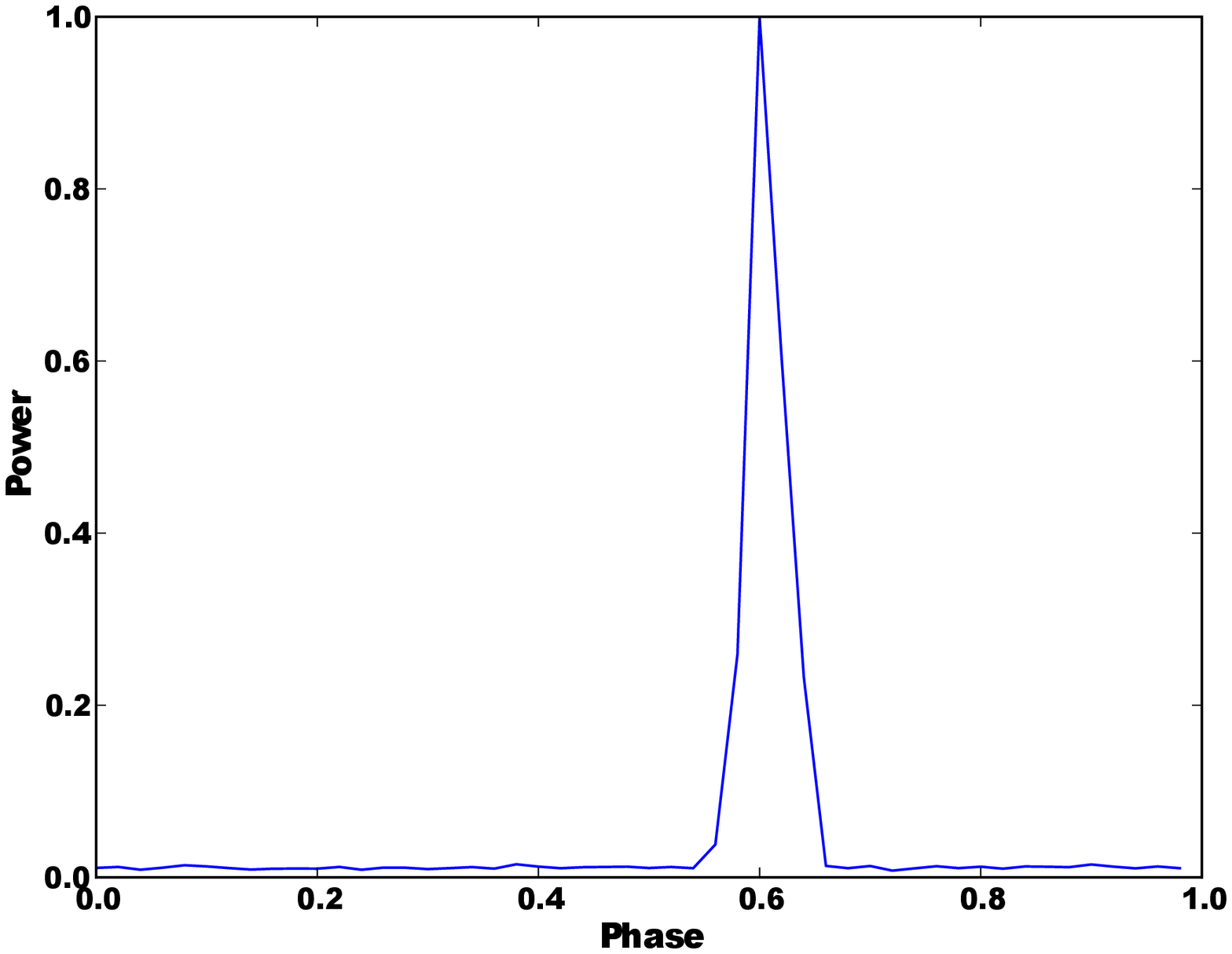}}
 \subfloat[(b)][]{ \includegraphics[width=0.45\linewidth]{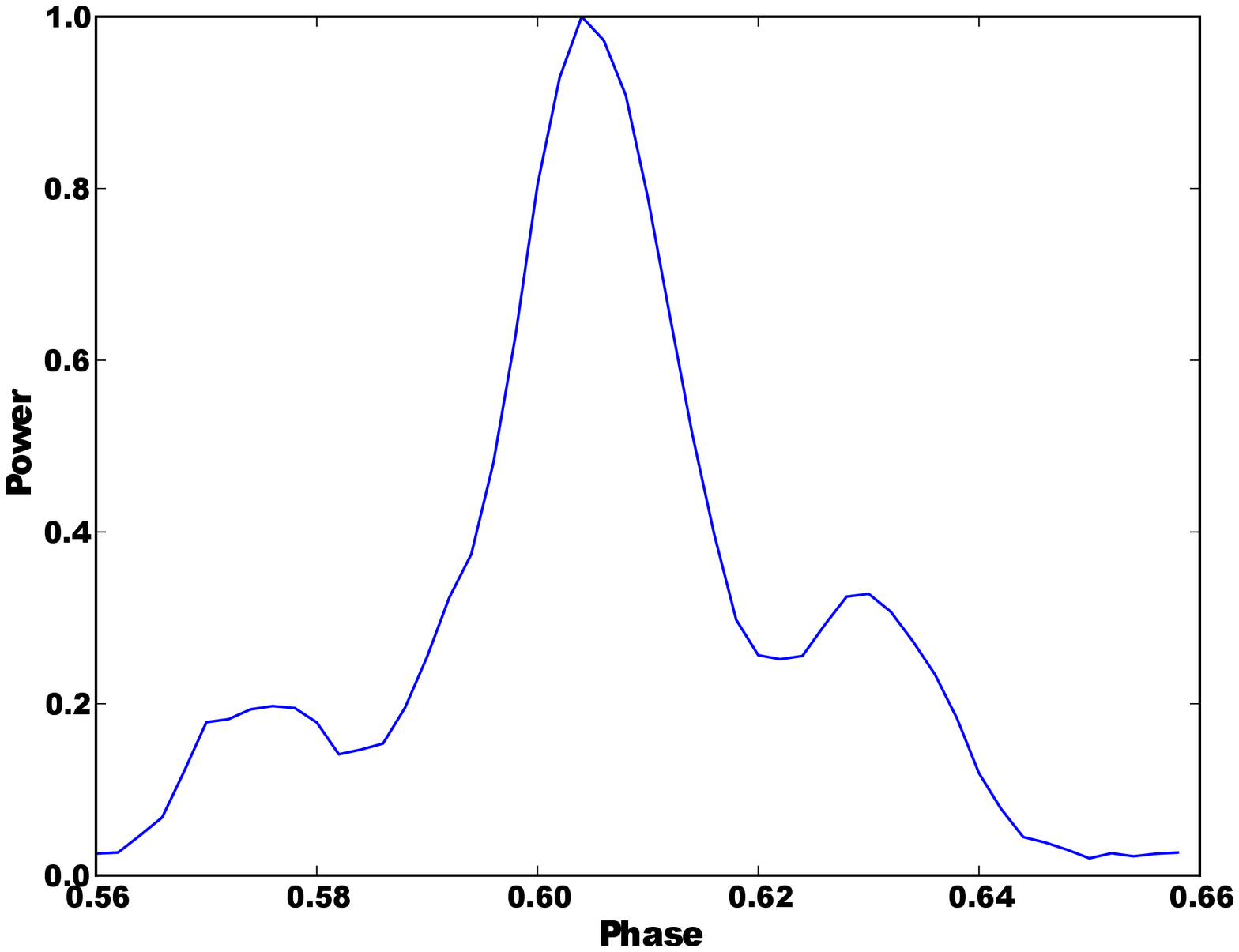}}
 \caption{\textbf{(a)} Pulse profile for B0329+54 obtained from 50 bins across the
          full period.
          The profile is obtained from the total power in each bin on the Effelsberg-Westerbork
          baseline. \textbf{(b)} same as (a) but with 50 bins across the pulse.}
 \label{fig:profiles}
\end{figure}

A closely related concept is pulsar binning. Here the pulse period is divided into a 
number of time bins and the correlation function is accumulated individually for each bin. 
In SFXC the user specifies a gating interval together with the
number of bins to be placed equidistantly within this interval.
In this implementation pulsar gating is merely a special case of pulsar binning. In
Fig.~\ref{fig:profiles} we show an example pulse profile of pulsar B0329+54 which was 
produced using the pulsar binning mode.
Pulsar binning requires that a model of the target pulsar is available, and therefore that it has
already been studied in a pulsar timing experiment. For SFXC the pulsar model is supplied in the 
form of a tempo(2) polyco file~\citep{Hobbs06}. The polyco file contains a polynomial model of the
pulsar phase as a function of time for a certain site and for a single frequency. SFXC requires the
polyco file to be generated for the geocentre.

An important consideration in pulsar observations is the dispersion caused by the interstellar
medium. The difference in arrival times (in ms) between two frequencies $\nu_1$ and $\nu_2$ 
(in MHz) is~\citep{Lorimer05} 
\begin{equation} 
  \Delta{t} \approx 4.15 \times 10^6 \times DM \times(\nu_1^{-2}-\nu_2^{-2})\quad[\textrm{ms}],
  \label{eq:pb-dt} 
\end{equation} 
where $DM$ is the dispersion measure. The dispersion measure is defined as the integrated electron
density along the line of sight to the pulsar. For some millisecond pulsars this dispersive
effect can be significant even within a single frequency band. 
For example the difference between the arrival times of the highest and lowest frequency
across an 8~MHz band at 21~cm for pulsar B1937+21 is 1.7~ms, which is 
larger than the pulse period\footnote{B1937+21 has a rotational period
$P = 1.56\mathrm{ms}$ and $DM = 71\mathrm{cm}^{-3}\mathrm{pc}$~\citep{Cognard95}}.

A computationally inexpensive method to de-disperse a signal is incoherent de-dispersion. Because the
dispersive delay is equal for all telescopes, incoherent de-dispersion can be applied post
correlation. The cross-correlations are performed at high enough spectral resolution such
that within a single frequency channel the dispersive delay is not significant. Each spectral 
point is then added to the correct pulsar bin by time shifting it according to 
Eq.~\eqref{eq:pb-dt}. This scheme works as long as the length of the FFTs required to reach the
needed spectral resolution is smaller than the pulse width. 

The dispersive delay can be removed completely using coherent de-dispersion
algorithms~\citep{Hankins75}. This involves applying a filter with transfer function 
\begin{equation}
H(\nu_0 + \nu) = \exp{\left(\frac{-i2\pi \nu^2DM}{2.41\times10^{-10}\nu_0^2(\nu_0+\nu)}\right)}
\label{eq:coh-dedisp}
\end{equation}
where $\nu_0$ is the central frequency of the observing band in MHz, and $|\nu|\le B/2$ where $B$
is the bandwidth. 

Applying a coherent de-dispersion filter comes at considerable computational cost compared to
incoherent methods. For the majority of pulsars the incoherent de-dispersion scheme is
sufficient. However, for some high-DM millisecond pulsars, coherent de-dispersion is necessary 
as the size of the FFTs that would be needed to perform incoherent de-dispersion
is close to the pulse period. At the time of writing (early 2014), coherent de-dispersion in 
SFXC was still in the validation phase
%
\section{Wide-field VLBI}
%
Fundamentally, the field of view in a VLBI observation is limited only by the primary antenna beams
of the participating telescopes. Provided that sufficient spectral and temporal resolution is
available to keep smearing effects at an acceptable level, it is possible to image the entire primary
antenna beam. However, this has been done only for
a small number of experiments
in the past, because of the very large data volumes~\cite[see e.g.][]{Lenc08, Chi13}.

A simple "first order" model of a radio dish is that of a uniformly illuminated
circular aperture of diameter $d$. The primary beam power $I(\theta)$, as a function of the angle 
$\theta$ (in radians), is determined by the Fraunhofer diffraction pattern of the aperture and
is given by~\citep{BornWolf}
\begin{equation}
I(\theta) = \left(\frac{2J_1(\pi d\sin(\theta)/\lambda)}{\pi d\sin(\theta)/\lambda}\right)^2,
\label{eq:wf_I}
\end{equation}
where $\lambda$ is the wavelength and $J_1(x)$ a Bessel function of the first kind.
From Eq.~\eqref{eq:wf_I} it follows that a circular aperture has its half maximum 
at $\theta=0.51\lambda/d$ and the first null at $\theta=1.22\lambda/d$.
In practice the circular aperture is only an approximation of a real physical dish. The 
beam shape ultimately has to be determined experimentally and in general is a function of
frequency and elevation. 

For an interferometer consisting of identical antennas, such as the VLBA, the beam
size of the interferometer is equal to the beam size of the individual antennas. For a
heterogeneous array such as the EVN the situation is more complicated~\citep{Strom04}. 
For example, consider the limiting case where one antenna is much larger than the
other dishes in the array and the beams of the smaller antennas can be considered constant
over the beam of the larger dish. In this case the size of the interferometer beam is up to
40\% larger than the beam of the largest dish in the array~\citep{Strom04}.

Besides primary beam losses, discretization effects in the correlator also are a source of signal
degradation. Two closely related effects are time- and frequency smearing. Both effects are ultimately
caused by the fact that the delay model is evaluated for a single point on the sky, the delay
tracking centre, but is then applied to the entire field. Time smearing occurs due to the fact 
that during each integration the correlation function is accumulated over some finite time
$T$. Because of the high delay rates in VLBI, contributions from sources which are not close to the
delay tracking centre will degrade quickly with $T$. Similarly, frequency smearing is caused by the
fact that each spectral channel can be regarded as an average over some bandwidth $\Delta\nu$. 
If the visibility phase varies significantly over this bandwidth then decorrelation will occur.
An in-depth review of time- and frequency smearing can be found in chapter 6 of~\citet{Thompson01},
and chapters 17 (W.D. Cotton) and 18 (A.H. Bridle and F.R. Schwab) in ~\citet{Taylor99}.

Because of the need for short integration times and high spectral resolution, conventional 
wide-field data sets can be very large.  For example, to map the primary beam of a 100m telescope 
at 21cm (9.5 arcmin) with a maximum baseline of 10000 km would
require 100ms integrations and 7.8 KHz channel widths to keep smearing
losses within 10\%. A standard 10 telescope 1 Gb/s EVN observation in this case would produce 
approximately 3 TB worth of user data in 6 hours. However, in an average field there will only be a 
few dozen mJy-class sources within the beam of a 100m dish. A more practical approach is to 
produce a separate narrow field data set for each source in the field
rather than a single monolithic wide field data set. This procedure greatly simplifies the data
reduction and is known as multiple simultaneous phase centre observing~\citep{Deller11}.

Internally the correlator processes the data at the required high spectral and temporal resolution,
performing short sub-integrations. At the end of each sub-integration, the phase centre
is shifted to all points of interest. The visibilities for each point of interest are accumulated
individually. The results are averaged down in time and frequency before they are written to disk.

The phase centre is shifted from the original phase centre to the target position
by applying a phase shift $\phi$ proportional to the difference in geometric delay $\Delta\tau$
between both positions, where~\citep{Morgan11}
\begin{equation}
\Delta\tau=(\tau'-\tau)(1-\dot{\tau}),
\label{eq:wf_dtau}
\end{equation}
here $\tau$ is the delay at the original phase centre, $\tau'$ the delay at the target position, and
$\dot{\tau}$ the time derivative of $\tau$. The term $(1-\dot{\tau})$ accounts for the change in
delay during the period $\tau-\tau'$. The phase shift $\phi$ then becomes
\begin{equation}
\phi=2\pi\nu_{\mathrm{sky}}\Delta\tau,
\label{eq:wf_phase}
\end{equation}
where $\nu_{\mathrm{sky}}$ is the sky frequency. 
\begin{figure}
 \centering
 \subfloat[(a)][]{ \includegraphics[width=0.49\linewidth]{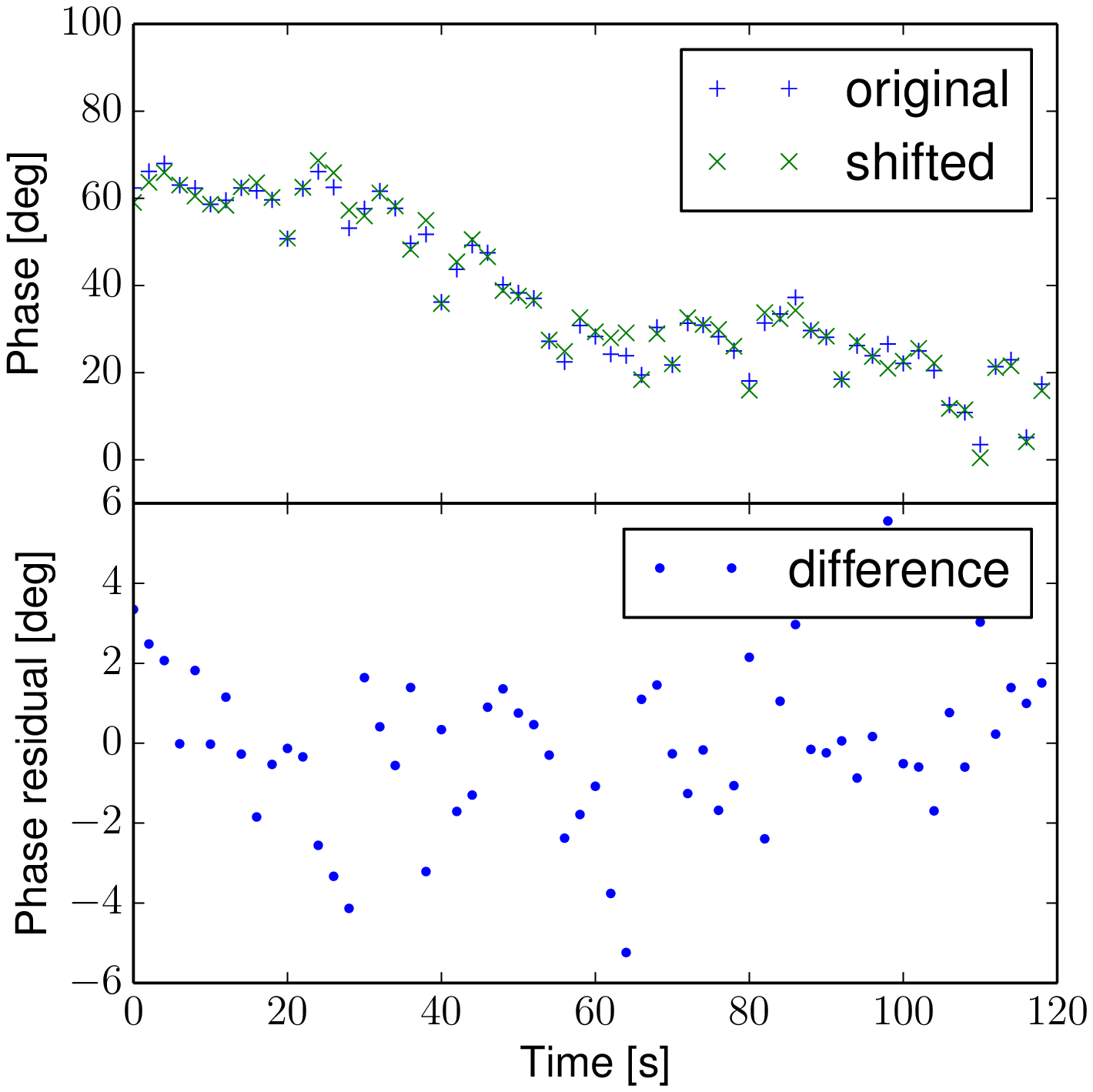}}
 \subfloat[(b)][]{ \includegraphics[width=0.49\linewidth]{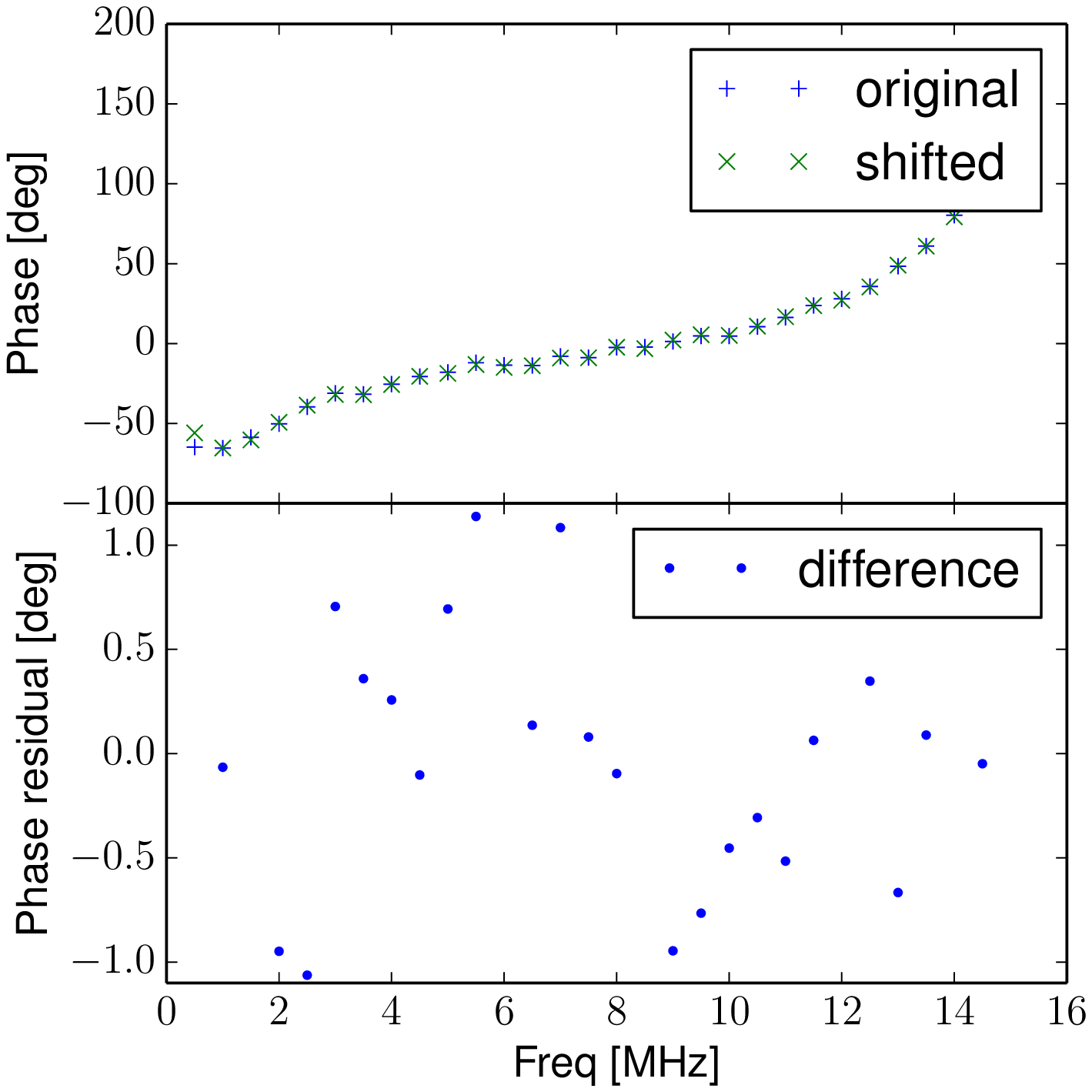}}
 \caption{Comparison between visibility phases obtained using the
  multiple simultaneous phase centre method and visibilities obtained using a conventional correlation
  on the Effelsberg - Shanghai baseline
 \textbf{(a)} Top: phase as a function of time for both datasets. Bottom: Phase difference between both datasets. The visibilities were averaged in frequency over the inner 80\% of the band.
 \textbf{(b)} Phase as a function of frequency, vector averaged over 120 sec.}
 \label{fig:wf_verfication}
\end{figure}
\begin{figure}
	\centering
	\includegraphics[width=0.5\linewidth]{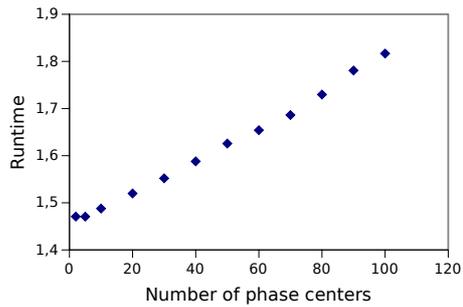}
	\caption{Performance of multiple phase centre correlation normalized to a single phase centre.
	}
	\label{fig:mphase-performance}
\end{figure}

We demonstrate the validity of the multiple phase centre method by comparing the following two 
data sets obtained using target source 3C66A. The first data set was created using a conventional
correlation with the delay tracking centre on the target. The second data set was obtained by placing the
primary delay tracking centre 3 arcmin away from the target source and obtaining a data set for 3C66A
using the multiple phase centre method. In both cases the same raw data were used and the 
pointing centre of the array coincided with 3C66A. The raw data consists of six 15 minute scans
that were distributed evenly over a 6 hour period. For the multiple phase centre
correlation, 8192 spectral points and sub-integrations of 25ms were used, keeping smearing
effects within 1\%. The end result was averaged down to 32 spectral points and 2s 
integrations. 
In Fig.~\ref{fig:wf_verfication} we show visibility phase for both datasets for a segment of data 
on the Effelsberg - Shanghai baseline. Both data sets track the same average phase and the phase
difference between both data sets is purely stochastic. Over the whole 6 hour data set we found the 
average phase difference to be $<\phi_1-\phi_2>=-0.01$~deg and the standard deviation of the phase
difference to be $\sigma_\phi=1.7$~deg. This is consistent with the shifted data set having a 
slightly lower signal to noise but does not imply a significant amount of decorrelation. 

Because the phase centre shifting is only performed at the end of each sub-integration period, 
which is relatively infrequently, the performance scales very well with the number of phase centres.
In Fig.~\ref{fig:mphase-performance} we plot the correlation time as 
function of the number of phase centres. The benchmark was performed using 10 minutes of data at
18~cm using 10 telescopes and a maximum baseline size of 10000 km. Internally 4096 spectral
points and 50ms sub-integrations were used, giving a total smearing loss of about 2\%.
Enabling multiple phase centre correlation increases correlation time by nearly 50\%, which is
mostly due to the large FFT size needed for this mode. After paying this constant performance
penalty, each additional phase centre comes at very little cost. Increasing the number of phase
centres from 2 to 100 only increases correlation time by 25\%.
%
\section{Software Architecture} 
%
\begin{figure} 
  \centering 
  \includegraphics[width=0.8\linewidth]{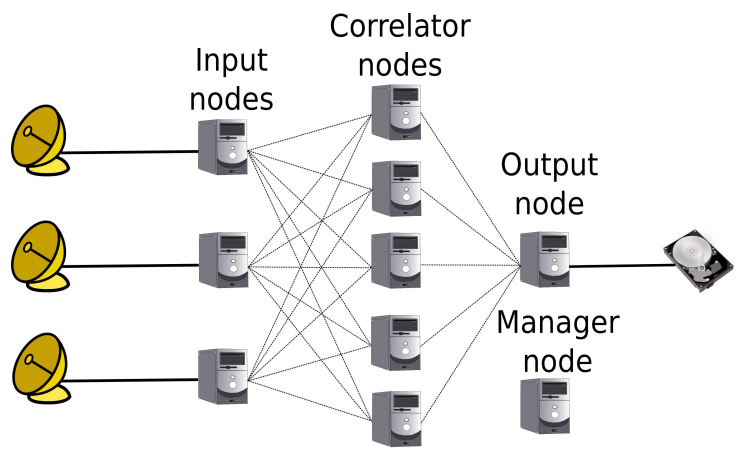}
  \caption{Overview of the SFXC software architecture.}
  \label{Fig:architecture}
\end{figure}

SFXC is a distributed MPI~\citep{MPI} application intended to be deployed on a generic Linux
computing cluster. The correlator is driven by two configuration files, a VLBI EXperiment
(VEX\footnote{\url{http://www.vlbi.org/vex/}}) file and a Correlator Control File (CCF). The VEX 
file describes the observation as it has been recorded, including the frequency set-up, 
the scan structure, clock offsets, etc. The VEX files are used by both the
telescopes and the correlator. In addition a CCF is supplied to the correlator which includes all
parameters specific to a correlation job. These parameters include information such as integration times,
data locations, and number of spectral channels. The CCF file is in
JSON\footnote{\url{http://json.org}} format which has the benefit of being human readable 
as well as being easy to parse in software.

In Fig.~\ref{Fig:architecture} we show the architecture of SFXC. The MPI processes are divided over
a number of specialized nodes, each performing a specific task. There are four types of nodes:
input nodes; correlator nodes; one output node; and one manager node. 
%
\subsection{Manager node}
%
The manager node controls the correlation process. The input data stream is divided into time
slices with a duration equal to the integration time. The main task of the manager
node is to assign time slices to correlator nodes. To increase the amount of parallelization each time
slice contains data only for a single sub-band. Thus each sub-band is sent to a different correlator
node.

Communication between the manager node and the other nodes is done via MPI messages. 
Whenever a correlator node has nearly finished correlating\footnote{Currently this is at 90\%} 
it will signal the manager node. The manager node will then assign a new time
slice to that correlator node so that the correlator node can immediately process a new
time slice when it finishes the current time slice.
 The relevant parameters pertaining to this time slice are 
passed to the input nodes and the correlator node itself.

When all time slices are processed the manager node will instruct all other nodes to shut down
and close the application.
%
\subsection{Input nodes}
\label{sec:inputnode}
\begin{figure}
  \includegraphics[width=0.5\linewidth]{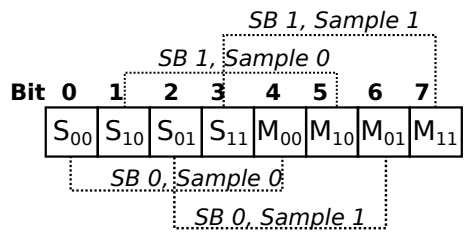}
 \caption{Example of a multiplexed 8 bits long data word as encountered in disk based data formats.
  In this example there are two sub-bands, called SB 0 and SB 1. Each sample consists of two bits, a Sign bit
  and a Magnitude bit. It is the task of the corner turner to split such a word into separate
  data streams for each sub-band.
  }
 \label{Fig:mplexdata}
\end{figure} 
An input node reads data from a single data source, corner turns the data into individual 
sub-bands, and sends the resulting data streams to the correlator nodes. In this process it applies
the integer delay correction, as described in Sec.~\ref{sec:coralg}.
For performance reasons, the data transfers are not done through MPI but via standard TCP
connections. The data source can be either a file, a network socket, or a Mark5
diskpack~\citep{Whitney03}. Currently the Mark4, Mark5B, VLBA, and VDIF dataformats are supported.

Most data formats are multiplexed. This is an artefact from the time when
data were recorded onto magnetic tapes. The tape was divided into tracks and each sub-band was 
split over one or more tracks. In principle the way these sub-bands are divided over the tracks 
is arbitrary and the exact mapping is recorded in the VEX file.
The concept of a track is converted to a disk based format by packing all tracks into a stream 
of input words. The size of each data word in bits is equal to the number of tracks, with 
a logical one-to-one mapping of bit location inside a data word to a track number. In this scheme 
the first track is mapped to bit 0, and so on, as illustrated in Fig.\ref{Fig:mplexdata}. 
Currently only the VDIF format~\citep{Whitney09} offers the possibility to record each sub-band 
in a separate data frame and therefore avoid the need for corner turning, although VDIF supports
multiplexing as well. 

Because the corner turning involves many bitwise operations it is a very costly in terms of
processing power. Due to the high data rates in VLBI, typically 1 Gb/s, the corner turning can
easily become a performance bottleneck. The corner turner consists of series of bitwise shifts,
and because the track layouts are in principle arbitrary the size of these shifts are not known
until run-time. However, it was found that the performance of the corner turner roughly doubles if 
all these parameters are known at compile time. To take advantage of this fact the corner turner 
is implemented as a dynamically loaded shared object which is compiled on the fly at run time with
all parameters supplied as constants. Furthermore, multiple 
corner turner threads are launched to maximize performance.

%
\subsection{Correlator nodes}
%
Each correlator node processes one time slice of a single sub-band from all telescopes. If 
cross-hand polarization products are required both polarisations are processed on the same 
correlator node. 

The data received from the input node are quantized at either 1- or 2 bits per sample. 
The data stream is first expanded to single precision 
floating point. After floating point conversion the remaining delay compensation and cross-correlation
steps described in Sec.~\ref{sec:coralg} are performed, with the exception of the integer delay
compensation which is performed in the input nodes. 
 
Typically the number of correlator nodes is chosen to be equal to the number of free CPU cores 
in the cluster. In Sec.~\ref{sec:performance} we will discuss the performance of the correlation
algorithm further.
%
\subsection{Output node}
\label{sec:output_node}
All data are sent to a single output node which aggregates all correlated data and
writes the result to disk in the correct order. The data is output in a custom data format which
can be converted to an CASA (AIPS++ version 2.0) measurement set. This measurement set in turn
can be converted to the FITS IDI format using existing tools developed for the Mark4 correlator.
%
\section{Validation}
\label{sec:validation}
%
\begin{figure}
\centering
\subfloat[(a)][]{\includegraphics[width=0.45\linewidth]{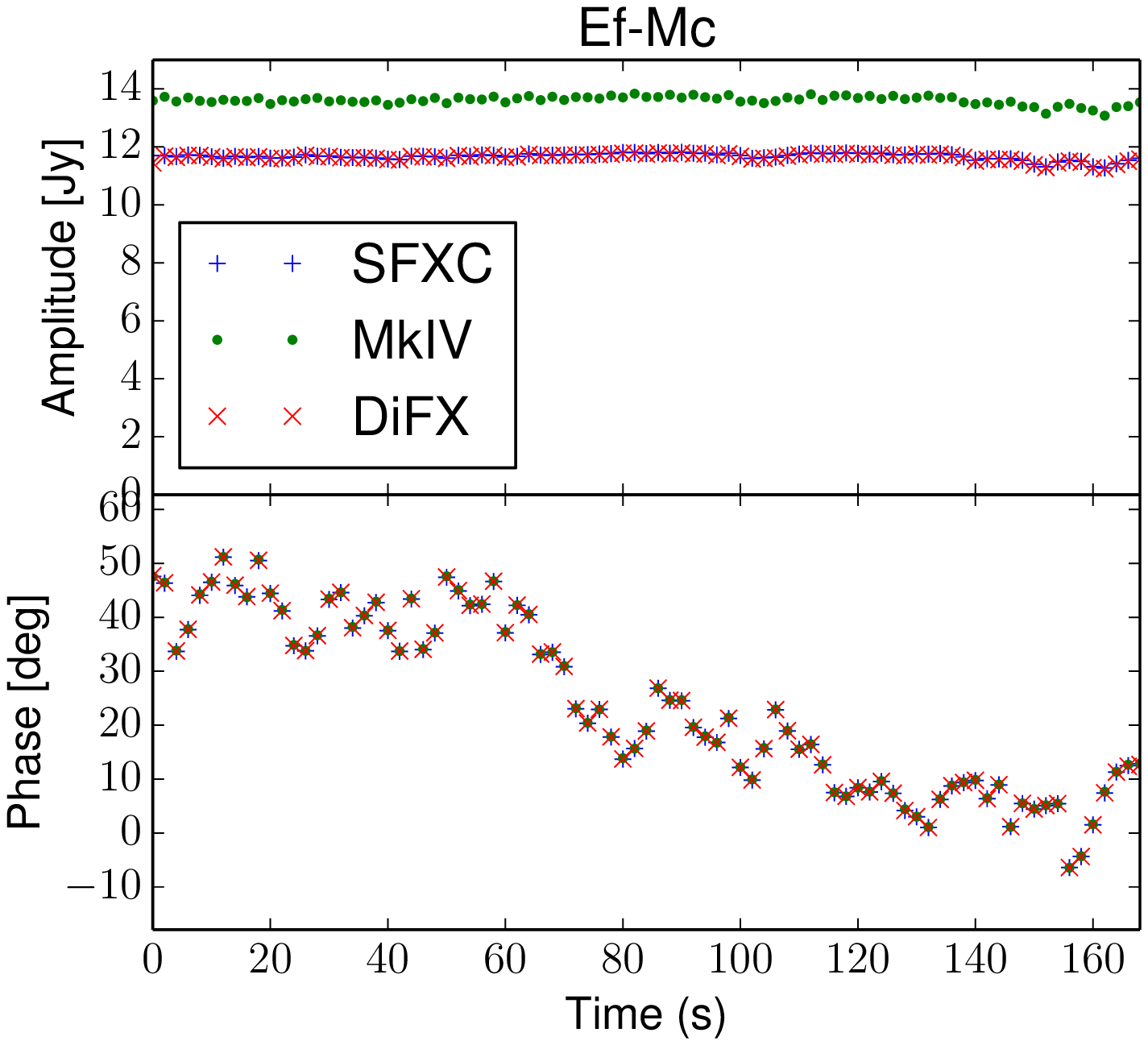}}
\subfloat[(b)][]{\includegraphics[width=0.45\linewidth]{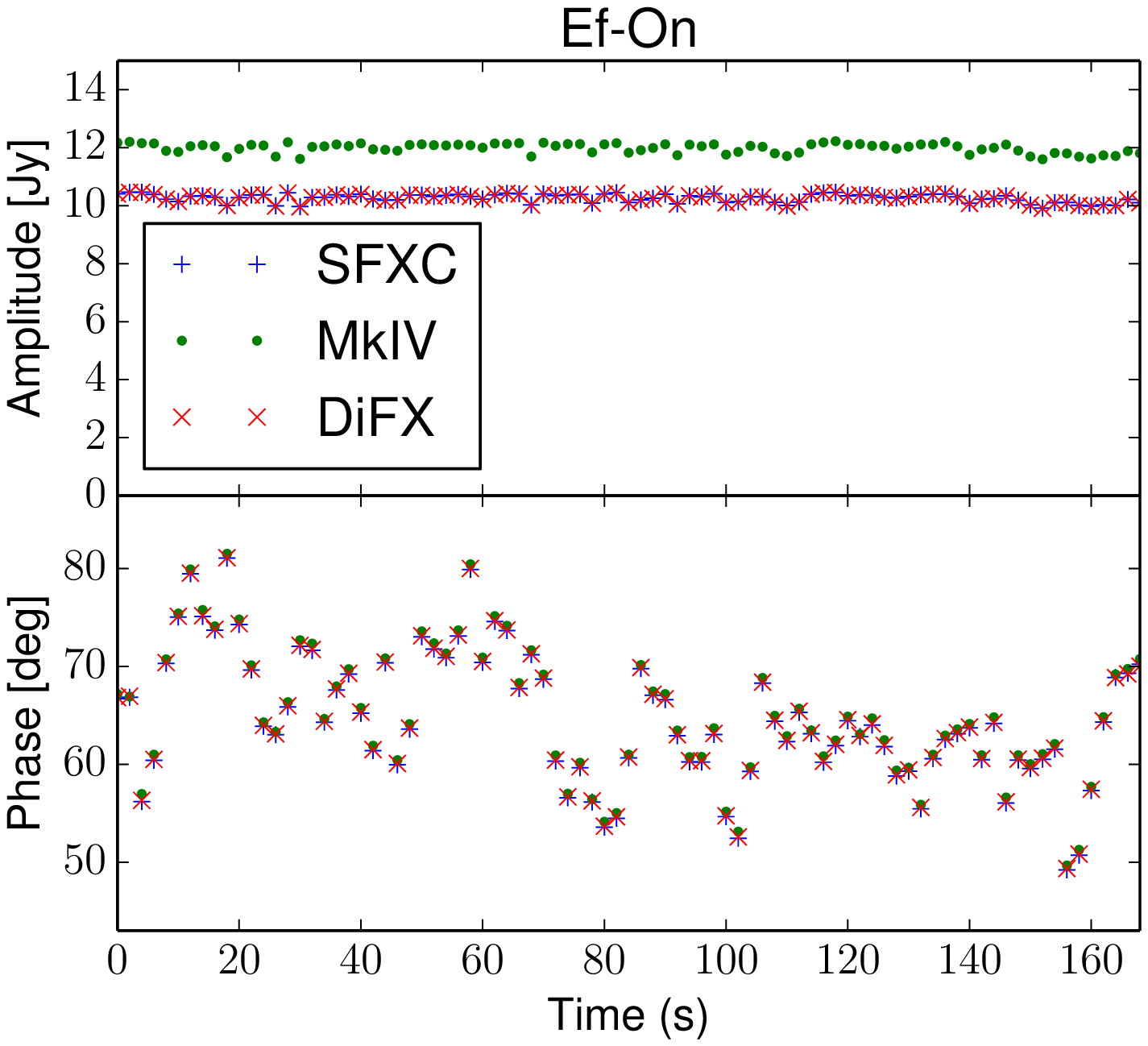}}
\newline
\subfloat[(c)][]{\includegraphics[width=0.45\linewidth]{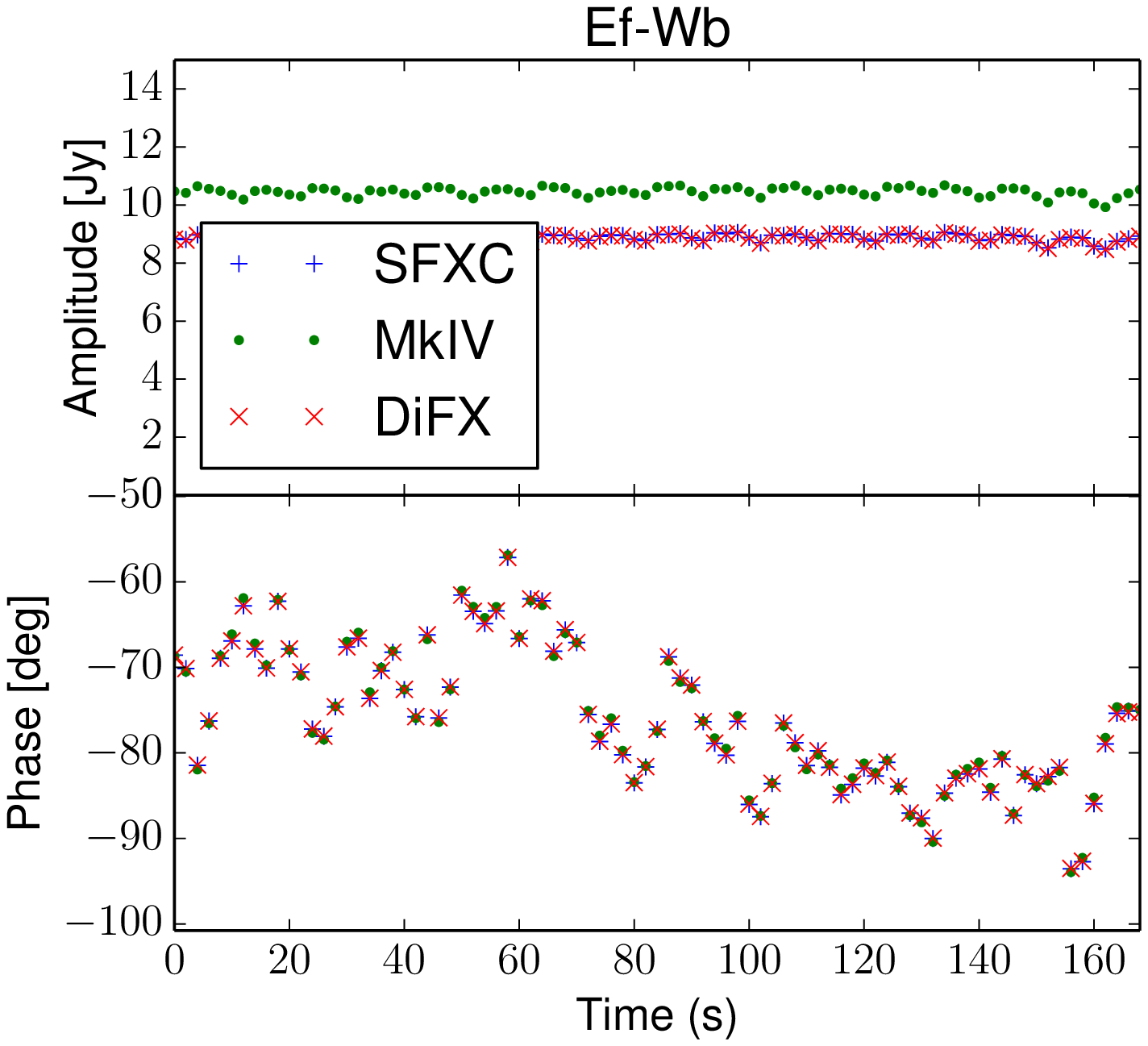}}
\subfloat[(d)][]{\includegraphics[width=0.45\linewidth]{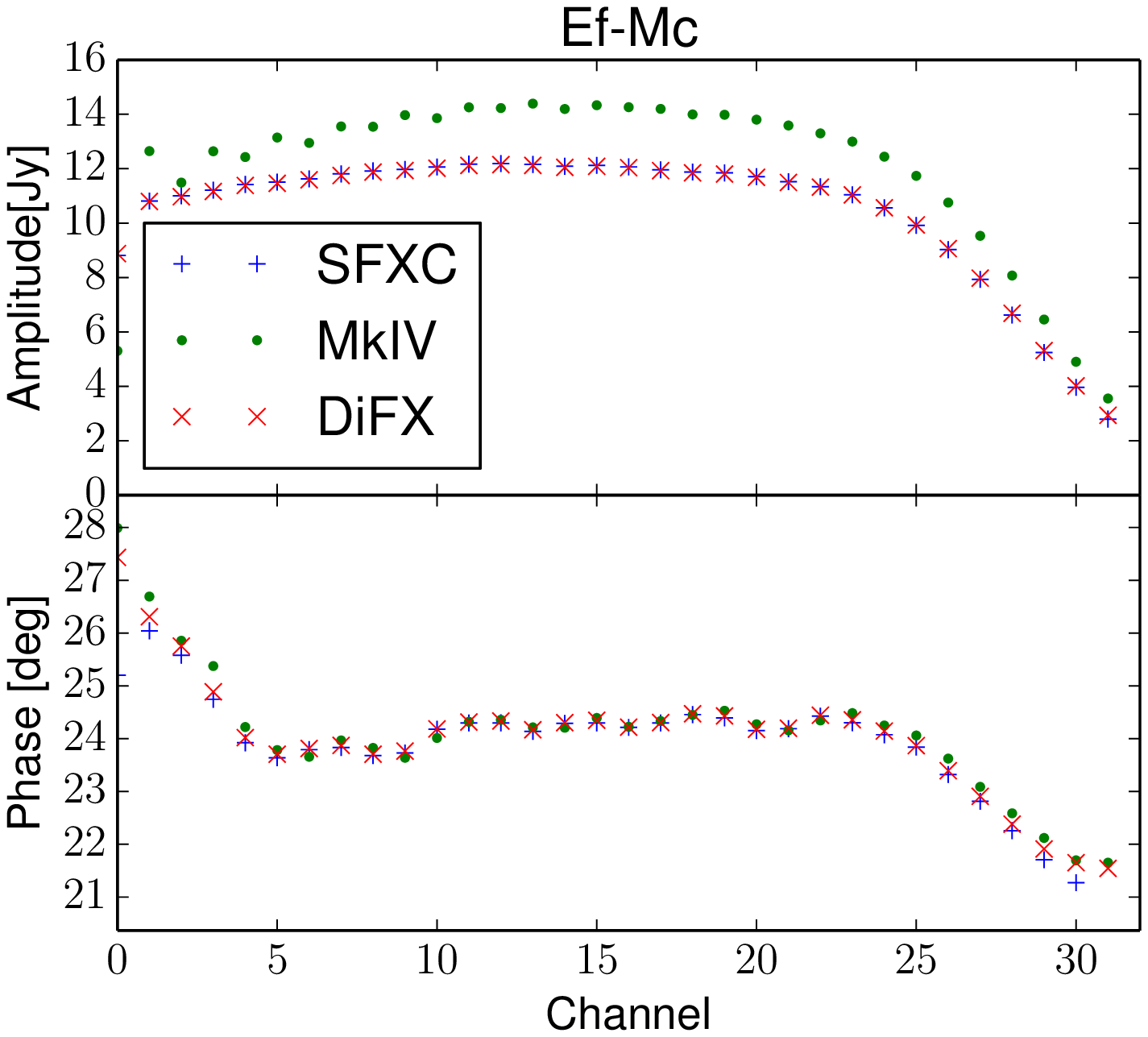}}
\newline
\subfloat[(e)][]{\includegraphics[width=0.45\linewidth]{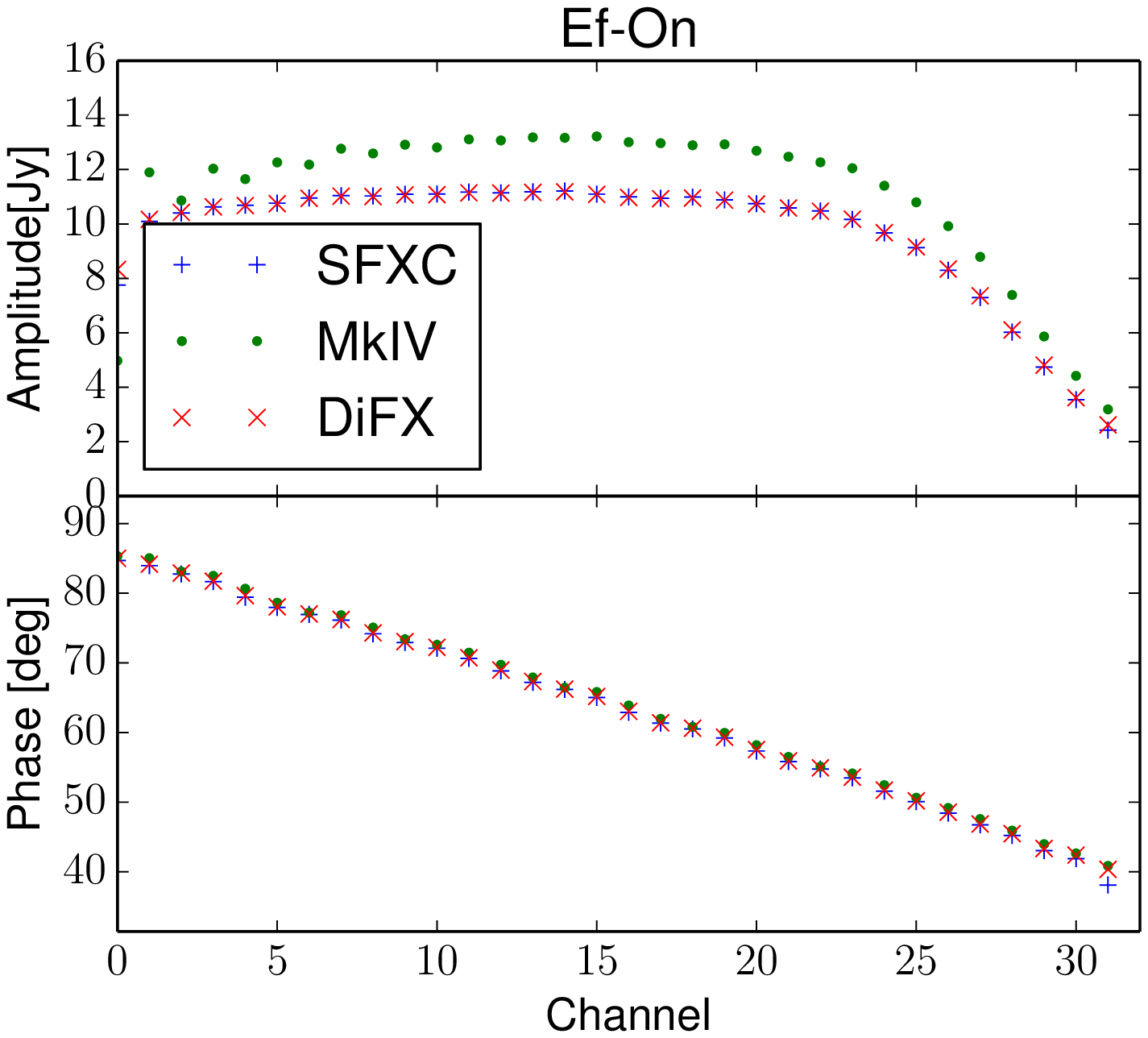}}
\subfloat[(f)][]{\includegraphics[width=0.45\linewidth]{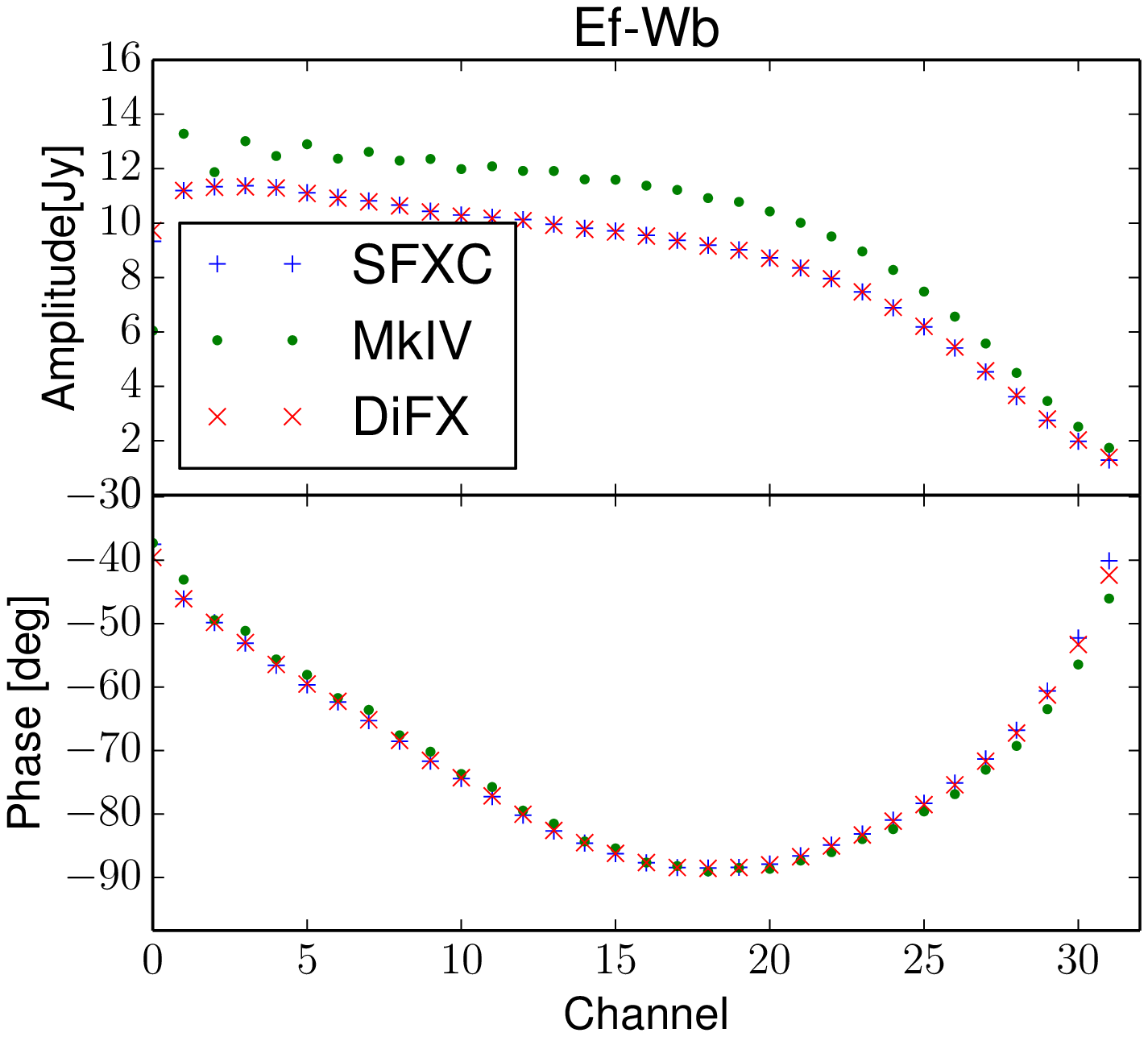}}
\caption{Comparison of visibility phase and amplitude between 
SFXC, DiFX, and the EVN MkIV data processor at JIVE for baselines from Effelsberg(Ef) to
Medicina(Mc), Onsala(On), and Westerbork(Wb). For details about the observation we refer to the main text.
\textbf{(a)}-\textbf{(c)} Amplitude and phase as a function
of time, averaging in frequency over the inner 80\% of the band. \textbf{(d)}-\textbf{(f)}
Amplitude and phase as a function frequency, averaging in time over 170 seconds.
}
\label{Fig:validation}
\end{figure}
In this section we give a brief comparison between SFXC, DiFX, and the EVN MkIV data 
processor at JIVE. To this end we correlated an observation of quasar 4C39.25 at 6~cm 
wavelength using these three correlators. The observation was performed using 8~MHz sub-bands and 
dual circular polarizations. The data was quantized using two bit sampling and was recorded in the 
mark4 data format, which all three correlators can read natively.

By default DiFX uses a different delay model than SFXC and the EVN MkIV data
processor at JIVE. To make a direct comparison of visibility phases possible we overrode 
the default delay model in DiFX and used the same CALC 10 delay model with all three correlators.

As was noted in Sec.~\ref{sec:output_node}, SFXC and the MkIV data processor share the same post-
correlation toolchain~\citep{Campbell08}, which applies a number of corrections to the data. 
For SFXC only a quantization loss correction is applied to the data. The fact that the input signal is
quantized using a finite number of bits leads to a lowering of visibility amplitudes~\citep[Sec 8.3]{Thompson01}. 
For ideally quantized two bit data this effect leads to a reduction in amplitude by 12\%.
There are a number of additional corrections that are applied to the data produced by the MkIV data
processor, which compensate for various approximations that are made
in the MkIV data processor's correlation algorithm~\citep{Campbell08}. These corrections scale visibility
amplitudes by an additional 22\%.

The data was correlated using 32 spectral points per sub-band, and a 2 second integration time.
The raw correlated data was converted to FITS-IDI and loaded into the data reduction package 
AIPS~\footnote{\url{http://www.aips.nrao.edu/}}. Using the AIPS task ACCOR a quantization loss correction was applied
to the DiFX data. As was noted previously, both SFXC and the MkIV already had this correction 
applied to their data.

In Fig.~\ref{Fig:validation} we show visibility phase and amplitude for a number of baselines. 
The visibility phases from the three correlators agree to a fraction of a degree of phase with each other.
Unfortunately, the EVN MkIV data processor at JIVE suffered from known amplitude scaling issues.
In Tab.~\ref{tab:comparison-ampl} we show the ratios between the visibility amplitudes from
the three correlators. While the amplitudes from both SFXC and DiFX agree very closely, the amplitudes 
from the MkIV data processor are on average 17\% higher than the other two correlators.
Similar amplitude scaling issues were reported for the MkIV correlator at Haystack
observatory~\cite{Cappallo11}. It should be emphasized that these higher correlation amplitudes are 
purely a scaling issue and do not correspond to a higher signal to noise.
\begin{table}
\centering
\begin{tabular}{|l|l|l|l|}
\hline
Baseline & SFXC / MkIV & SFXC / DiFX & DiFX / MkIV\\
\hline
Ef-Mc    & 0.857 & 1.002 & 0.856\\
Ef-On    & 0.856 & 1.000 & 0.856\\
Ef-Wb    & 0.849 & 1.002 & 0.847\\
\hline
\end{tabular}
\caption{Visibility amplitude ratios for the data shown in Fig.~\ref{Fig:validation}.}
\label{tab:comparison-ampl}
\end{table}%
\section{Performance} 
\label{sec:performance}
%
\begin{figure}
	\subfloat[(a)][]{\includegraphics[width=0.48\linewidth]{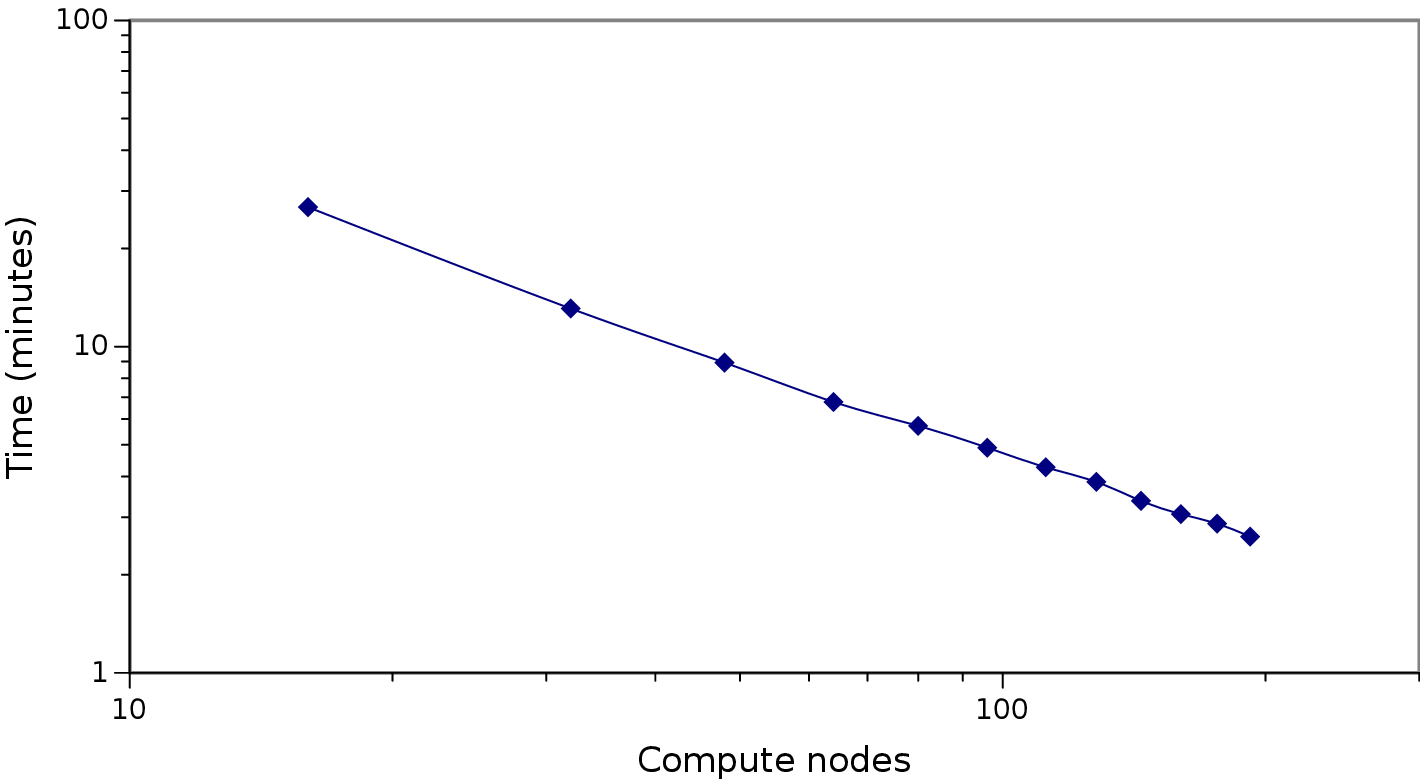}}
	\subfloat[(b)][]{\includegraphics[width=0.48\linewidth]{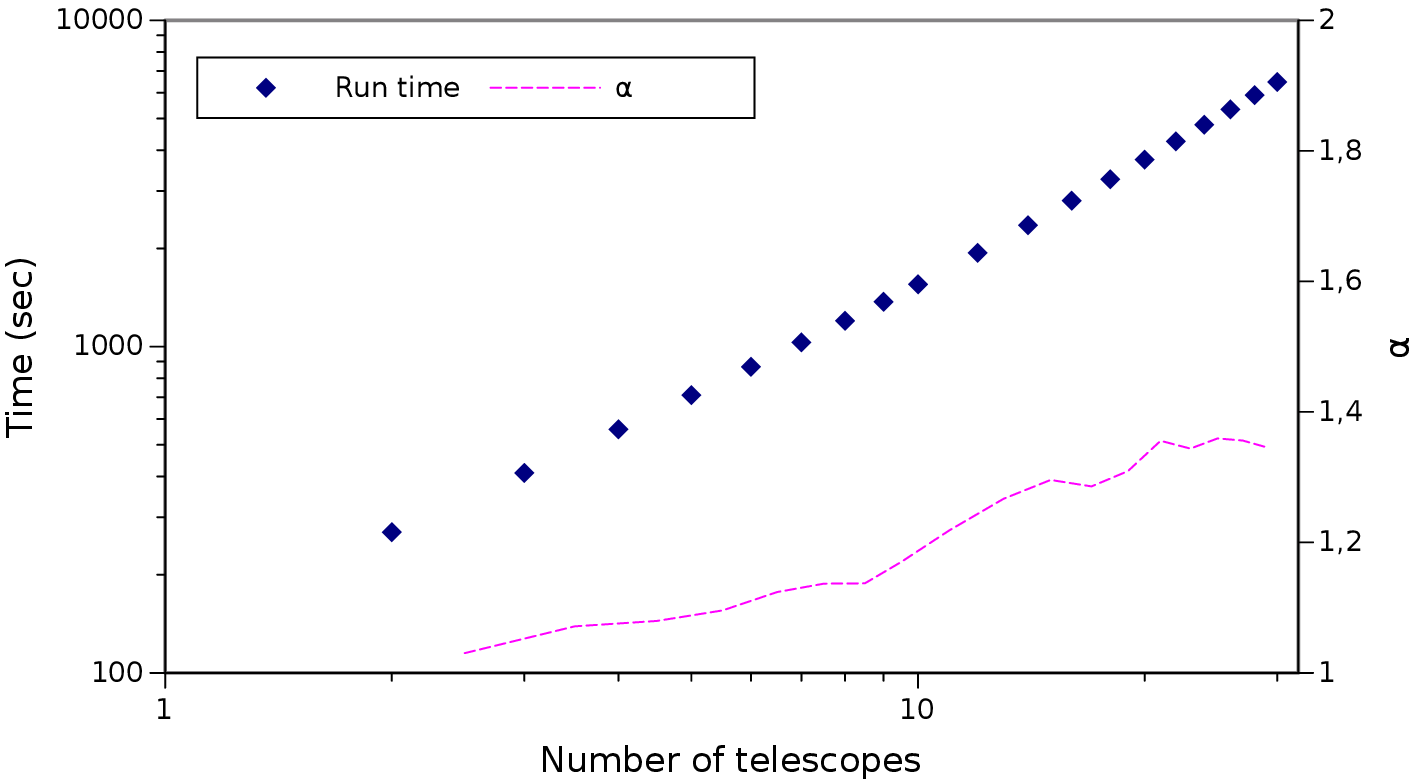}}
	\caption{(a) Correlation time as a function of the number of correlator nodes,
	         (b) Correlation time $T$ as a function of the number of telescopes $N_t$ together with
	         the exponential growth factor $\alpha$ which is computed by fitting the relation
	        $T=C N_t^\alpha$ to consecutive data points, where $C$ is a constant.
	        }
	\label{fig:performance-benchmark}
\end{figure}
In this section we present a number of benchmarks performed on the EVN software correlator cluster 
at JIVE. The cluster consists (2014) of 40 nodes all of which are equipped with two Intel Xeon processors,
yielding a total of 384 CPU cores. The cluster nodes are interconnected via QDR 
Infiniband, while  each node is connected to the outside world via two bonded gigabit Ethernet
links. In addition a number of nodes are equipped with 10 Gb/s Ethernet interfaces to accommodate 
e-VLBI. Currently the cluster is capable of processing 14 stations at 1 Gb/s datarate in real-time.

When available, SFXC will utilise the Intel Performance Primitives (IPP) library. 
This library provides vectorized math routines 
for array manipulation utilizing the SIMD instructions available 
in modern Intel-compatible CPU's. Furthermore, the library includes highly optimized FFT routines.
If the IPP library is not present SFXC will fall back on the FFTW3~\citep{FFTW05} FFT library and
uses its own non-SIMD routines for array manipulation. Benchmarks on recent Intel Xeon processors
show a performance gain of typically 40\% using the IPP libraries. 
The SFXC installed on the EVN software correlator cluster at JIVE makes use of the IPP libraries.

Software correlation is a data intensive, embarrassingly parallel process and therefore highly
scalable. Performance often is determined by the rate at which data can be supplied to the 
correlator nodes. As we discussed in Sec.~\ref{sec:inputnode} the input nodes 
perform a highly CPU-intensive corner turning task. Therefore care should be taken that sufficient
resources are available to the input nodes.

Provided that enough IO capacity is available, the performance of the correlator
scales linearly with the number of CPU cores. This is illustrated in 
Fig.~\ref{fig:performance-benchmark}-(a) where we show the correlation time as a function of the
number of correlator nodes.  The test was performed on 10 minutes of data recorded at 512 Mb/s 
with 10 telescopes. 

Computational cost in terms of the number of telescopes can be divided into two regimes. In 
the first regime
antenna-based operations, such as delay compensation, dominate. In this regime the correlation
time increases approximately linearly with the number of telescopes. In the second regime
baseline-based operations dominate, and there the correlation time increases quadratically with
the number  of telescope. Two kinds of baseline-based operations have to be performed, namely 
cross-correlations and phase shifts during multiple phase centre correlations. Because these phase
shifts are applied relatively infrequently, the cross-correlations dominate the processing.  
In Fig.~\ref{fig:performance-benchmark}-(b) we show correlation time as a function of the number of
telescopes.  The test was done using 20 minutes of data at 512 Mb/s.
The number of correlator nodes was limited to 32 which prevents the computation becoming IO limited
at the lower bound. As can be seen, the relation between correlation time and the number of 
telescopes only very slowly departs from linearity, and even for 30 telescopes the growth factor is
still below 1.5.

\begin{figure}
  \centering
  \includegraphics[width=0.7\linewidth]{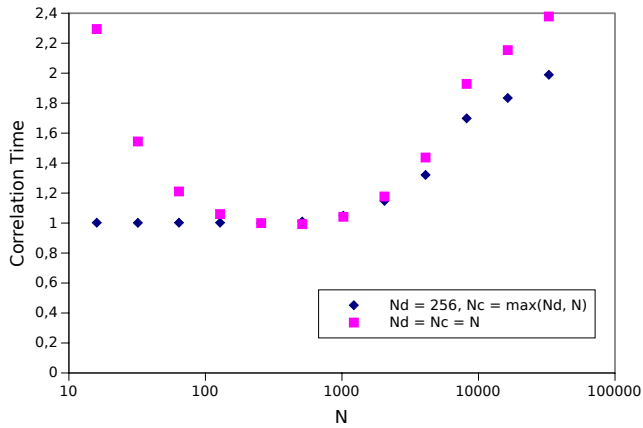}
  \caption{Correlation time as a function of the number of spectral points in the output data $N$ for two configurations of the number of frequency points in the delay compensations $N_d$ and the number of points in the correlation $N_c$. The configuration signalled by the blue diamonds is the SFXC default.
  }
  \label{fig:bench-fft}
\end{figure}
As we described in Sec.~\ref{sec:coralg} the number of spectral points in the delay compensation
$N_d$ and the number of spectral points in the cross-correlations $N_c$ are independent of each
other. In Fig.\ref{fig:bench-fft} we show the correlation time as a function of $N_c$ for the
case when $N_c=N_d$, and for the case when $N_d=256$ for all $N_c$, the latter case being
the default. This benchmark shows that the optimum $N_d$ equals $256$ or $512$ for our computing
hardware. Both $N_c$ and $N_d$ are user configurable but are not mandatory for the user
to specify. When the user requests $N$ spectral points in the output data the following
defaults for $N_d$ and $N_c$ will be chosen: 
\begin{itemize}
\item When $N\le256$, both $N_d$ and $N_c$ are set to 256. The correlation products are 
averaged down to $N$ points after integration.
\item When $N>256$, $N_d$ is set to 256, and $N_c$ is equal to $N$.
\end{itemize}
%
\section{Conclusions and discussion} 
%

During the past years the SFXC software correlator has been deployed
as the operational EVN correlator at JIVE. As such it has proven to be
a highly reliable and stable platform.

We argue that software correlators in general have a number of
important advantages over their hardware counterparts. Software is
very flexible and easily modified, making it possible to add new
functionality with much less effort. This is exemplified by the
ever-increasing set of capabilities in SFXC, such as arbitrary
spectral resolution, spectral windowing, pulsar binning and gating,
including (in)coherent de-dispersion, and correlation with multiple
simultaneous phase centres. None of these features were available on
the MkIV correlator, nor would they have been straightforward (if at
all possible) to implement. This demonstrates how the implementation
of a software correlator has brought new science capabilities to the
EVN.

Another advantage is that software correlators are far more flexible
with respect to their input parameters, while hardware correlators as
a rule impose hard limits on parameters such as number of spectral
points or number of input stations. Once built, it is nearly
impossible to increase the size of a hardware correlator, as most
components are custom-made and produced in limited quantities. On the
other hand, as shown in Sec. 6, the performance of SFXC scales very
well with the available computing resources. As a consequence it is
nearly trivial to increase the capacity of a software correlator, only
involving the purchase of additional off-the-shelf hardware. And as a
result of Moore's law, these additions will be more powerful
and consume less power, at a lower price.

\begin{acknowledgements}
Part of this work was made possible by the SCARiE project under the NWO STARE programme, and by
the EC-funded EXPReS and NEXPReS projects, project numbers 026642 and RI-261525.
GC and DAD acknowledge the EC FP7 project ESPaCE (grant agreement 263466).
DAD and GMC acknowledge the EC FP7 project EuroPlaNet (grant agreement 228319).
GMC acknowledges also the NWO-ShAO agreement on collaboration in VLBI.
We thank Max Avruch and Ruud Oerlemans for their contributions
during the early stages of SFXC development.

\end{acknowledgements}
\bibliographystyle{spbasic} \bibliography{sfxc}{}
\end{document}